\documentstyle[prb,aps,epsfig,multicol]{revtex}

\begin{document}

\title{Metastable states of a driven flux lattice in a superconductor with strong pins}
\author{L. Fruchter}
\address{Laboratoire de Physique des Solides,
Universit\'{e} Paris-Sud, C.N.R.S., B\^{a}t. 510, 91405 Orsay
France}

\date{\today}
\maketitle

\begin{abstract}
The flux lattice driven by a uniform driving force in a
superconductor with hot, strong, sharp and randomly distributed
pinning centers, with applied magnetic field half the matching
field is simulated. At low temperature both a non activated
regime, where flux motion occurs within a robust percolative flux
flow channel, and an activated regime are obtained depending on
the sample preparation. These two regimes exhibit distinct
resistivity and magnetic induction. In the non activated regime,
a clear fingerprint is observed in the autocorrelation function
of the longitudinal resitivity, which oscillates at a frequency
close to the inverse lattice diffusion time.
\end{abstract}

\pacs{PACS numbers: 74.60.Ge}

\begin{multicols}{2}
\narrowtext

\section{Introduction}

High-temperature superconductors have revealed a complex static
vortex lattice phase diagram in the temperature-magnetic field
(or disorder strength) plane, and it is now well established, both
theoretically and experimentally, that at least three distinct
phases are present : the high temperature liquid phase, the
quasi-ordered and the glassy solid phases at low temperature for
low and high magnetic field
respectively\cite{feigel89,houghton89,giam97}. Introducing a
driving force as a third axis, leads to the even more complex
dynamic phase diagram of a driven vortex lattice. A theoretical
insight on this situation has been given in refs.
\onlinecite{nattermann90,shi91,koshelev94,balents95,giam96,ledoussal98}.
The emerging picture is the existence, for increasing drive, of
creep, plastic and moving solid regimes\cite{yaron95}. The moving
solid, once thought a 'moving crystal' within the perturbation
theory\cite{koshelev94}, was found to be a 'moving glass' in which
vortices flow along elastically coupled
channels\cite{giam96,ledoussal98}. Such a theoretical work has
widely benefited from the comparison with numerical experiments.
Simulations first demonstrated the existence of a plastic flow
regime within channels for the highly defective
lattice\cite{brandt83,jensen88,brass89}, yielding non-linear I-V
characteristics\cite{shi91,bhatta94}. Dynamic phase transitions
or crossovers between the plastic regime and moving solids for
lattices driven by a uniform external force were put into
evidence in refs
\onlinecite{koshelev94,moon96,ryu96,olson98,kolton99}. The case
of the field gradient driven lattice (Bean state) was extensively
studied by F. Nori et al (see e.g. ref. \onlinecite{olson98b} and
refs therein). Besides these effort to determine a univoque
dynamic phase diagram for the driven lattice, some evidence exist
for the existence of multiple metastable states. Indeed, the
static glass is characterized by the existence of infinite
barriers that prevent the exploration of the entire phase diagram
and the existence of a degenerated ground state. As a
consequence, one might also expect that the driven lattice can
show multiple metastable states close to the static glassy phase,
i.e. in the plastic regime. Such a behavior was observed in ref.
\onlinecite{gronbech96}, where it was shown at zero temperature
that some filamentary flow channels are stable in a finite range
of driving force. The transition between different flow channels
structures as the driving force is varied was shown to result in
steps in the $I-V$ curves. In the following, I show that a random
distribution of strong pinning centers allows for the realization
of metastable states of the disordered driven lattice at non zero
temperature.

\section{Experimental details}
\label{experiment}

A two dimensional lattice submitted to a uniform driving force in
the presence of 'hot pins' is simulated. A semi-infinite sample is
considered. Two opposite edges of the sample are submitted to the
external magnetic field, $B_{0}$, which is simulated by an extra
force $f_{B_{0}}$ acting on each vortex, perpendicular to the
sides of the sample. The force acting on a vortex at a distance
$x$ from the edge is the one imposed by a semi-infinite vortex
lattice at a distance $a_{0}+x$, where
$a_{0}=\left(\Phi_{0}/B\right)^{1/2}$ is the flux lattice spacing
at the equilibrium. Whenever the force acting on a vortex
situated at the sample edge is directed towards the inside of the
sample, this new vortex is introduced. The other two edges of the
sample are subject to a periodic boundary condition, thus
simulating a semi infinite geometry.

Flux lines are assumed rigid rods. This correctly modelizes a
layered superconductor with decoupled layers, or a three
dimensional material with a large line energy. The force per unit
length between vortices separated by a distance $r$
is\cite{brandt83a}:
\begin{equation}
f_{vv}(r)=\left(\Phi_0^{2}/8\pi^{2}\lambda^{3}\right)
K_{1}\left(r/\lambda\right)
\label{fvv}
\end{equation}
where $K_{1}$ is a Bessel function. This is a good approximation
strictly only in the case of vortex lines (rods) and for 2D
vortices a logarithmic interaction should be used. In the present
case, the more rapid decrease of the Bessel function allows us to
cut the interaction between vortices at a distance $5\lambda$ and
save computation time.

Strong pinning centers are randomly distributed in the sample.
The density of the pinning sites is $n=B_{\Phi}/\Phi_{0}$, with
$\Phi_{0}$ the flux quantum and $B_{\Phi}$ the 'matching field'
for which an equilibrium flux line lattice shows the density of
flux lines $n$. The pinning sites are assumed normal cylinders
parallel to the applied field, with radius $c_0$. The situation
where, at low temperature, the vortex core radius $\xi(T)$ is
smaller than $c_0$ and pinning is due essentially to the
reduction of the core energy when the line sits on the pin is
considered. The force per unit length exerted by a pin at a
distance $r$ to the line is given by:
\begin{equation}
f_{p}(r) = \sigma \varepsilon_{o} r/r_{0}\xi\;  \textrm{for $r
\leq r_{0}$}\;
 0  \;\textrm{for $r > r_{0}$}
\label{fp}
\end{equation}
where $\varepsilon_{0} = \left(\Phi_{0}/4\pi\lambda\right)^{2}$
is the line energy, $r_{0}=c_{0}+\xi/2$ and $\sigma \leq1$. The
pinning force is exactly balanced by the Lorentz force,
$\mathbf{J}\wedge\mathbf{\Phi_{0}}$, for $J = J_{p} = \sigma
\varepsilon_{o}/\xi\Phi_{0}$. From eq. \ref{fvv} and \ref{fp},
the energy scale for vortex repulsion is $U_{vv} = 2
\varepsilon_{0} K_{0}(a_{0}/\lambda)$ and the one for pins
attraction is $U_{p}=\sigma \varepsilon_{0} r_{0}/2\xi$.

The temperature in the whole sample is assumed zero, excepted on
the pinning sites ('hot pins' model). That is, thermally activated
depinning from the pinning centers is present, while thermal
fluctuations of the unpinned vortices are absent. Within such a
model, there is no melting of the flux line lattice, and the
contribution of the thermal fluctuations of the unpinned vortices
to depinning is neglected. During the simulation, the force
acting on the vortices, and the corresponding activation barrier
for the lines which are pinned, are computed. The equation of
motion:
\begin{equation}
\eta \;d\mathbf{r}/ dt =
\mathbf{f_{B_{0}}}+\mathbf{f_{vv}}+\mathbf{f_{p}}+
\mathbf{J}\wedge \mathbf{\phi_{0}}
\end{equation}
is applied for a time step $\Delta t$ small enough so that vortex
displacements within this interval are much smaller than $a_{0}$.
For the pinned vortices, escape from the wells proceeds according
to the Arrhenius exponential probability. During the simulation,
the vortices trajectories are registered, as well as the
resistivity, normalized to its flux flow value :
\begin{equation}
\rho / \rho_{F} = (J B_{0} \Delta t)^{-1}\;\sum_{i}\;
x_{i}(t+\Delta t)-x_{i}(t)
\end{equation}
In the rest, the driving current is normalized to $J_{p}$ and the
thermal energy $k_{B}T$ to the pinning energy $U_{p}$. A sample
of dimensions $40 a_{0}$ along the driving current and $30 a_{0}$
perpendicular to it was studied. The following superconducting
parameters were used: $\lambda = 1400 \;\AA$, $\xi = 18 \;\AA$,
$r_{0} = 2 \xi$, $\sigma = 0.1$, yielding $U_{p}/U_{vv} = 0.075$,
and $B_{\phi} = 2 B_{0} = 5000 \;$G. This situation, using the
terminology in ref.\onlinecite{brandt83a}, is the one of random,
\textit{sharp} ($r_{0} \ll a_{0}$) and \textit{strong}
($U_{p}/r_{0}c_{66} \gg 1$) pins close to the matching field.

\section{Results and Discussion}
\begin{figure}
\begin{center}
\epsfig{file=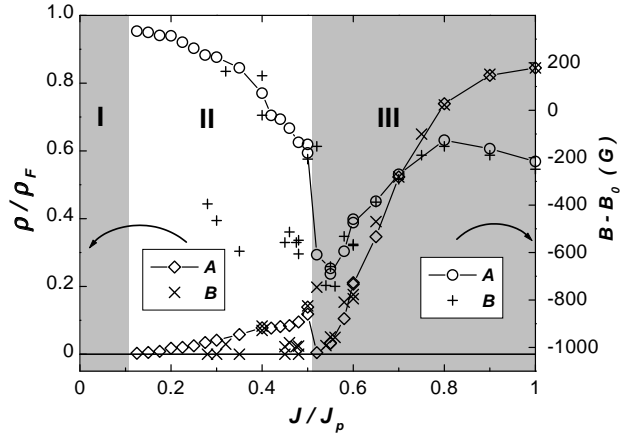, width=8.4cm}
\end{center}
\caption{Resistivity and magnetic induction for procedures {\bf
\textit{A}} and {\bf\textit{B}} as described in the text ($T =
0$).} \label{VI}
\end{figure}

I have first computed the resistivity and the induction at $T =
0$ using two different procedures (Fig.\ref{VI}). Both procedures
start from a regular hexagonal lattice at equilibrium with the
external magnetic field and $J = 0$. In the first procedure, the
driving current is first set to $J = J_{p}$. It is then decreased
by steps down to $J = 0$. A steady state - characterized by a
steady resistivity and induction - is allowed to settle at each
step. In the second procedure, the driving current is directly
set from zero the measuring value. I denote {\bf \textit{A}} and
{\bf \textit{B}} these two procedures, and $\rho_{A}$ and
$\rho_{B}$ the associated resistivity. From these experiments, it
is possible to distinguish three regimes. {\bf I}: for $J
\lesssim 0.1$, $\rho_{A} = 0$ and $\rho_{B}=0$. {\bf II}: for
$0.1 \lesssim J \lesssim 0.5$, $\rho_{A} \neq 0$ and $\rho_{B}$
is found randomly on two branches, one of them coinciding with
$\rho_{A}$, the other being zero or close to zero. {\bf III}: for
$0.5 \lesssim J$, both procedures yield similar non zero values.
The existence of a clear transition between regimes {\bf II} and
{\bf III} is also clearly put into evidence by procedure {\bf
\textit{A}} only, as both $\rho_{A}$ and the induction show a
marked discontinuity (Fig.\ref{VI}). In regime {\bf II}, the
existence of two branches for procedure {\bf \textit{B}} mirrors
in the induction which also splits into two distinct curves, one
of them coinciding with the one obtained with procedure {\bf
\textit{A}}. For all regimes, no Bragg peaks could be detected
from the time-average structure factor
%$S(\mathbf{k})=
%<\left(\Sigma_{i}\exp(i\;\mathbf{k}\;\mathbf{r}_{i})\right)^{2}>_{t}$
and flux lines ordering is amorphous. Vortices trajectories in
regimes {\bf II} and {\bf III} both consist in irregular channels,
characteristic of plastic
motion\cite{brandt83,jensen88,jensen88b}.

\begin{figure}
\begin{center}
\epsfig{file=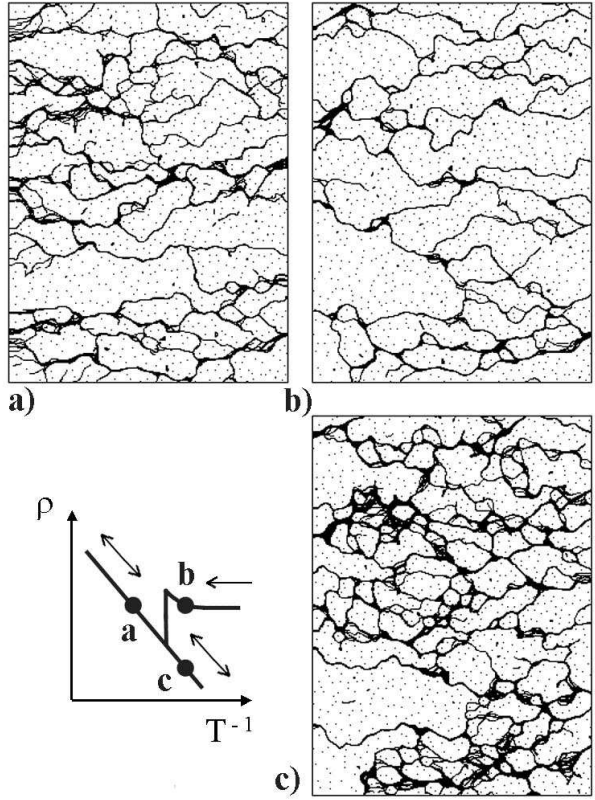, width=8.4cm}
\end{center}
\caption{Vortices trajectories ($J=0.2$) at $k_{B}T = 0.04$ ({\bf
a}) and $k_{B}T = 0.025$ ({\bf b}, {\bf c}). The resistivity is
the same for configurations {\bf a} and {\bf b}. During the
recording, 90 vortices have crossed the sample edge for all three
samples. For sample {\bf b}, motion within the channels is
non-activated.} \label{modes}
\end{figure}

\begin{figure}
\begin{center}
\epsfig{file=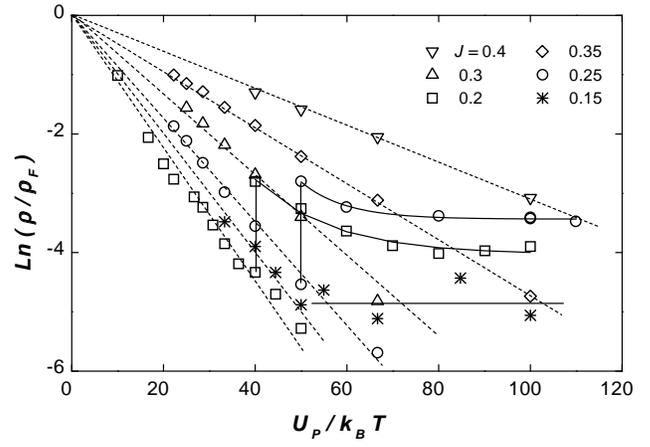, width=8.4cm}
\end{center}
\caption{Arrhenius plot of the resistivity upon heating from the
$\rho = 0$ branch in regime {\bf II} in Fig.\ref{VI} (dashed
line) and from the $\rho \neq 0$ one (full line). The dashed line
may be described for both increasing and decreasing temperature.}
\label{arrhenius}
\end{figure}

\begin{figure}
\begin{center}
\epsfig{file=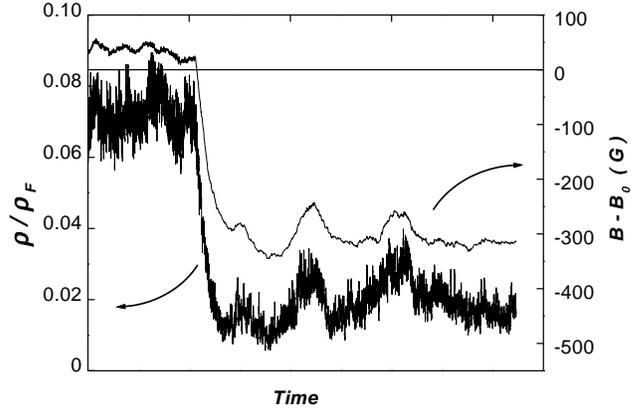, width=8.4cm}
\end{center}
\caption{Resistivity and induction at $k_{B}T = 0.03$ showing a
spontaneous switch from the unactivated regime to the activated
one. The sample has been progressively heated from the $\rho \neq
0$ branch in Fig.\ref{VI} ($J = 0.2$).} \label{transition}
\end{figure}

\begin{figure}
\begin{center}
\epsfig{file=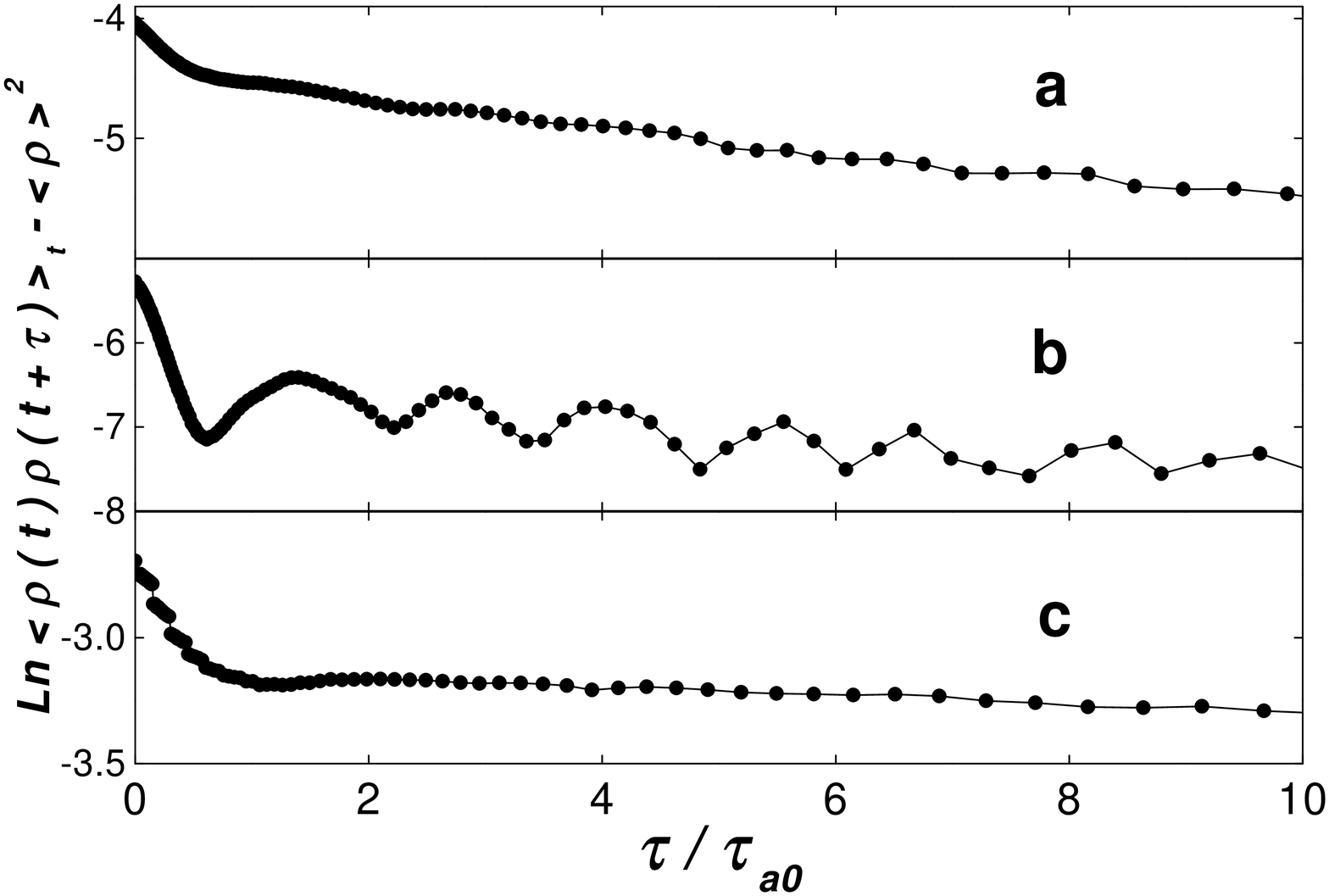, width=8.4cm}
\end{center}
\caption{Logarithm of the autocorrelation of the resistivity
noise for the three samples in Fig.\ref{modes}. The
autocorrelation function for the unactivated sample shows marked
oscillations at a frequency $\approx 0.7$ the inverse lattice
diffusion time.} \label{correlations}
\end{figure}

\begin{figure}
\begin{center}
\epsfig{file=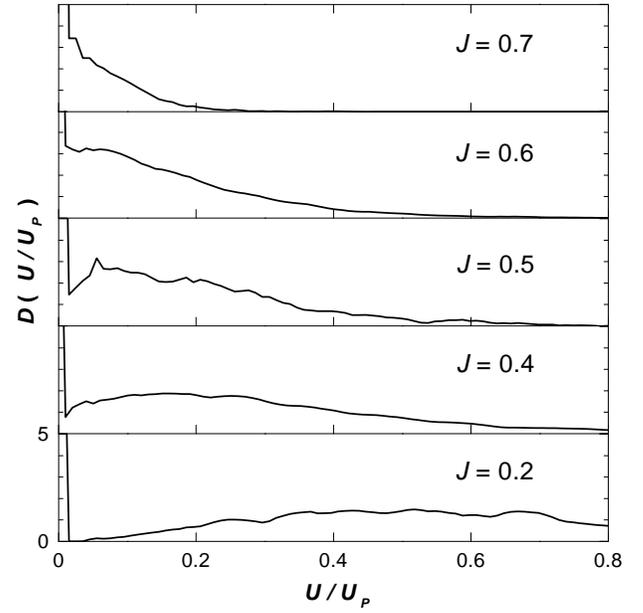, width=8.4cm}
\end{center}
\caption{Distribution of the activation energy, $U$, at $T = 0$
(normalized to area unity). Below $J \approx 0.5$, the
distribution for the pinned vortices shows a gap at low energy.}
\label{energies}
\end{figure}

I now argue that the upper branch of the resistivity in regime
{\bf II} is anomalous with respect to thermal activation for
depinning. Heating a sample with $0.1 < J < 0.5$ on the $\rho = 0$
branch from $T  = 0$ to finite values yields an exponentially
activated resistivity (Fig. \ref{arrhenius}). The activation
energy is found up to one order of magnitude lower than the one in
a one-dimensional model, $U(J) = (1 - J)^{2}$, which is due to the
inhomogeneous screening current in two dimensions. In a striking
different way, the upper branch of the resistivity is found
stable when temperature is increased, until the system switches
to the activated branch. Comparison of the vortices trajectories
for the activated regime and the non-activated one shows little
difference (Fig. \ref{modes}), but the autocorrelation function
of the resistivity noise, $<\rho(t) \rho(t+\tau)>_{\;t} -
<\rho>^{2}$ , exhibits oscillations at a frequency close to the
inverse lattice diffusion time, $1/\tau_{a0} = \phi_{0} J /c \;
\eta\; a_{0}$, in the case of the non-activated regime
(Fig.\ref{correlations}). This is the signature of the coherent
motion of vortices within flux flow channels, whereas activated
motion tends to randomize flux line motion. The distribution of
the activation energy for single vortex depinning, as shown in
Fig.\ref{energies} (we are here only concerned by individual
depinning, inherent to our thermal model), provides further
indication that the upper branch of the resistivity could be
stable against thermal depinning. It is clear from this data
that, as one enters regime {\bf II}, the distribution splits into
a dirac function at $U = 0$ for the unpinned vortices, and a wide
component that peaks at a non zero value with a gap at low
energy. It is then appealing to credit the observed stability at
low temperature of the pinned 'edifice' to the existence of this
gap. On the other hand, I have observed that two different runs
where the sample is heated from the $\rho \neq 0$ branch can
yield two slightly different temperatures for the switch to the
activated regime. Also, as can be seen in Fig. \ref{transition},
the system may spontaneously switch from the apparently stable
non-activated regime to the activated one at a constant
temperature. Whether the flux flow channels are truly stable upon
thermal excitations or there is an increase beyond the
experimental window of some relaxation time is thus an open
question.

Clearly, the existence of regimes {\bf II} and {\bf III} is
related to the possibility for the flux lattice to store elastic
energy (proportional to $B-B_{0}$ in Fig.\ref{VI}) in the
compression ({\bf II}, $B > B_{0}$) or the extension mode ({\bf
III}, $B < B_{0}$), as the transition between these two regimes
coincides with the one between these modes (Fig.\ref{VI}). In the
same way, in regime {\bf II}, the unactivated branch involves the
compression mode, whereas the activated one involves the
extension mode, as in regime {\bf III}. This is also clearly seen
in Fig.\ref{transition}, where bursts in the resistivity after
the system has switched to the activated mode correlate with the
ones of the induction. Then, both the existence of two distinct
regimes and of the anomalous unactivated branch for the
resistivity are a consequence of the applied field being close to
the matching field: for lower ratio $B/B_{\phi}$, I expect regime
{\bf II} to shrink.

Finally, the characteristic frequency for correlated motion
within flux flow channels may be evaluated from the conventional
Bardeen-Stephen expression for flux flow resistivity, yielding
$1/\tau_{a0}=J \rho_{n}c/a_{0} H_{c2}$. I consider the case of
2\textit{H}-NbSe$_{2}$ which is a good candidate for the
experimental realization of the above situation, as it exibits
low depinning critical current for as-grown crystals, as well as
a small line energy. We first need to evaluate the depinning
current at the transition between regimes {\bf I} and {\bf II},
$J_{I/II}$. At this point, the driving force on the flux flow
channel is balanced by the shear force exerted on it by the
pinned lattice : $J_{I/II}\; \Phi_{0} \approx c_{66}\; a_{0}$,
i.e. $J_{I/II} \approx \varepsilon_{0} / 4 \Phi_{0}\;a_{0}$. Using
parameters in section \ref{experiment}, we obtain $J_{I/II} =
0.05 \;J_{p}$, in reasonable agreement with the experimental data
(Fig. \ref{VI}). Going back to the case of
2\textit{H}-NbSe$_{2}$, using $H_{c2}= 2 \;10^{4}$ Oe, $\lambda =
2000 \;\AA \;$ and $\rho_{n} = 5 \;10^{-6}\; \Omega$ cm for the
normal state resistivity\cite{bhattacharya94,henderson96}, one
obtains $J_{I/II} \approx 2 \; 10^{3} B^{1/2}$ A cm$^{ -2}$ and
$1/\tau_{a0} \approx 90\;B$ kHz. For induction larger than a few
$10^{3}$ G, $J_{I/II}$ is well above the critical current density
usually observed for as-grown crystals\cite{henderson96}, so that
pinning by natural defects should not prevent the observation of
flux motion within the channels. However, the characteristic
frequency strongly exceeds the bandwidth of conventional four
probes transport measurements.

In conclusion, it was shown that a random distribution of pinning
centers with applied field close to the matching field allows for
the preparation of two competitive regimes of the disordered
driven flux lattice, within a large range of driving force. One
of them involves non activated flux flow motion within channels,
which are found robust against moderated thermal depinning. This
regime shows characteristic oscillations of the autocorrelation
function of the longitudinal resistivity at a frequency close to
the inverse lattice diffusion time.

\end{multicols}
\end{document}